\newcommand{\Teff}{T$_{\mathrm{eff}}$ }
\newcommand{\kms}{km~s$^{-1}$ }
\newcommand{\afe}{[$\alpha$/Fe] }
\documentclass[useAMS,usenatbib]{mn2e}
\usepackage{graphicx}
\usepackage{amssymb}
\usepackage{times}
\usepackage{subfigure}
\usepackage[T1]{fontenc}
\usepackage{aecompl}
\usepackage[breaklinks,colorlinks,citecolor=blue]{hyperref}

\begin{document}

\title[Characterizing High-Velocity Stars of RAVE]{Characterizing the High-Velocity Stars of RAVE: The Discovery of a Metal-Rich Halo Star Born in the Galactic Disk}
 \author[K.~Hawkins et. al.]{K.~Hawkins$^{1}$\thanks{E-mail: khawkins@ast.cam.ac.uk}, G.~Kordopatis$^{1, 2}$, G.~Gilmore$^{1}$, T.~Masseron$^{1}$, R.~F.~G.~Wyse$^{3}$, G.~Ruchti$^{4}$, \newauthor O.~Bienaym\'e$^{5}$, J.~Bland-Hawthorn$^{6}$, C.~Boeche$^{7}$, K.~Freeman$^{8}$, B.~K.~Gibson$^{9}$, E.~K.~Grebel$^{7}$, \newauthor A.~Helmi$^{10}$, A.~Kunder$^{2}$, U.~Munari$^{11}$, J.~F.~Navarro$^{12}$, Q.~A.~Parker$^{13,14}$, W. A. Reid$^{13}$, \newauthor R.~D.~Scholz$^{2}$, G. Seabroke$^{15}$, A.~Siebert$^{5}$, M.~Steinmetz$^{2}$, F.~Watson$^{14}$, T. Zwitter$^{16}$\\ 
$^{1}$Institute of Astronomy, Madingley Road, Cambridge CB3 0HA, UK \\
$^{2}$Leibniz-Institut f\"ur Astrophysik Potsdam (AIP) An der Sternwarte 16, D-14482 Potsdam, Germany \\
$^{3}$Physics and Astronomy Department, Johns Hopkins University, 3400 North Charles Street, Baltimore, MD 21218, USA \\ 
$^{4}$Lund Observatory, Department of Astronomy and Theoretical Physics, Box 43, SE-22100 Lund, Sweden\\ 
$^{5}$Observatoire astronomique de Strasbourg, Universit\'e de Strasbourg, CNRS, UMR 7550, 11 rue de l'Universit\'e, F-67000 Strasbourg, France\\
$^{6}$Sydney Institute for Astronomy, School of Physics A28, NSW 2006, Australia\\
$^{7}$Astronomisches Rechen-Institut, Zentrum f\"ur Astronomie der Universit\"at Heidelberg, M\"onchhofstr.\ 12-14, 69120 Heidelberg, Germany\\
$^{8}$Research School of Astronomy and Astrophysics, Australian National University, Cotter Rd., ACT 2611 Weston, Australia\\
$^{9}$Jeremiah Horrocks Institute, University of Central Lancashire, Preston, PR1 2HE, U.K.\\
$^{10}$Kapteyn Astronomical Institute, University of Groningen, PO Box 800, NL-9700 AV Groningen, the Netherlands\\
$^{11}$INAF Astronomical Observatory of Padova, 36012 Asiago (VI), Italy\\
$^{12}$CIfAR Senior Fellow, University of Victoria, Victoria, BC, Canada V8P 5C2\\
$^{13}$Research Centre for Astronomy, Astrophysics and Astrophotonics, Macquarie University, Sydney, NSW 2109, Australia\\
$^{14}$Australian Astronomical Observatory, 105 Delhi Road, North Ryde, PO Box 915, North Ryde, NSW 1670, Australia\\
$^{15}$Mullard Space Science Laboratory, University College London, Holmbury St Mary, Dorking, RH5 6NT, UK\\
$^{16}$Faculty of Mathematics and Physics, University of Ljubljana, Jadranska 19, SI-1000 Ljubljana, Slovenia\\ }


\date{Accepted 2014 December 02. Received 2014 December 01; in original form 2014 November 05}


\maketitle

\label{firstpage}

\begin{abstract}
We aim to characterize high-velocity (HiVel) stars in the solar vicinity both chemically and kinematically using the fourth data release of the RAdial Velocity Experiment (RAVE). We used a sample of 57 HiVel stars with Galactic rest-frame velocities larger than 275 km s$^{-1}$. With 6D position and velocity information, we integrated the orbits of the HiVel stars and found that, on average, they reach out to 13 kpc from the Galactic plane and have relatively eccentric orbits consistent with the Galactic halo. Using the stellar parameters and \afe estimates from RAVE, we found the metallicity distribution of the HiVel stars peak at [M/H] = --1.2 dex and is chemically consistent with the inner halo. There are a few notable exceptions that include a hypervelocity star (HVS) candidate, an extremely high-velocity bound halo star, and one star that is kinematically consistent with the halo but chemically consistent with the disk. High-resolution spectra were obtained for the metal-rich HiVel star candidate and the second highest velocity star in the sample. Using these high-resolution data, we report the discovery of a metal-rich halo star that has likely been dynamically ejected into the halo from the Galactic thick disk. This discovery could aid in explaining the assembly of the most metal-rich component of the Galactic halo.
\end{abstract}


\section{Introduction}
The advent of large spectroscopic surveys in the last decade has opened a new field of investigation: high-velocity (HiVel) stars. These stars are rare objects, defined by having velocities well above the typical speed of the stars one might expect \citep[e.g. $>$80 \kms relative to the Sun,][]{Schuster1988}, but below the Galactic escape speed. HiVel stars are intriguing in part because they can provide insight to the mechanism that produce their velocities. The origin of these HiVel stars can also provide useful information about the environments from which they are produced.  While recent studies have used only the kinematics of high-velocity stars to obtain an estimate of the Galaxy's mass \citep[e.g.][]{Smith2007, Piffl2013}, there have been only a few studies aimed at combining their chemical and kinematic information to get a picture of where these stars are produced and what caused them to achieve such high-velocities. Therefore, we aim to fill this gap by combining both the kinematics and chemistry of these HiVel stars using the RAdial Velocity Experiment \citep[RAVE]{Steinmetz2006} to discern if they are consistent with any particular component of the Milky Way and what may have produced them. 

\cite{Ryan2003} studied a sample of 10 intermediate-metallicity HiVel stars and found that most of them resemble the thick disk yet the HiVel stars of other studies \citep[e.g.][]{Schuster2006} suggested that HiVel stars are metal poor. This raises the question what is the chemical distribution (in [Fe/H] and [$\alpha$/Fe]\footnote{The $\alpha$-elements include those which have atomic numbers as a multiple of 4, such as Mg, Ti, Si, and Ca. [$\alpha$/Fe] in this paper is defined as the mean abundance of these $\alpha$-elements.} spaces) of HiVel stars in the solar neighbourhood? The answers to this question will ultimately aid in constraining where HiVel stars are born and thus help constrain models for how they are produced. In turn, this will help develop a better understanding for the assembly of the Galactic halo for which many of these HiVel stars are thought to currently reside. For example, if the metallicity distribution of HiVel stars is significantly more metal-rich compared to the halo, and if the \afe distribution is comparable to the disk, it may support the suggestion of \cite{Bromley2009} that the metal-rich tail of the Galactic halo may have come from kinematically heated, stars which formed in the disk.

These `runaway' disk stars described above are a subclass of HiVel stars and were first identified by \cite{Humason1947}. Runaway disk stars can provide an invaluable connection between star formation in the Galactic disk and halo. These stars are rare and described by peculiar velocities up to 200 \kms and are thought to have formed in the disk and ejected into the halo. Theoretically, runaway stars can be produced through a number of different mechanisms including: (1) binary supernova ejection \citep[e.g.][]{Blaauw1961, PortegiesZwart2000} and (2) dynamical ejection due to 3- and 4-body encounters \citep[e.g.][]{Poveda1967, Bromley2009}. It is thought that these above mechanisms can produce both low-mass and high-mass runaway stars. Yet most of the literature regarding runaway stars focuses on high-mass O and B type stars. This is likely because observationally O and B type runaway stars are bluer compared to normal, low-mass halo stars among where they were found \citep[e.g.][]{Poveda2005}. Characterizing HiVel stars, particularly in a data set with evolved low-mass stars, will allow us to search for these elusive stars. 

Another intriguing subclass of HiVel stars is hypervelocity stars (HVSs), which are racing through space at above the escape speed of the Milky Way. These stars are thought to be produced via three-body interactions between a binary star system and the super massive black hole at the Galactic Centre \citep{Hills1988}. However, other production mechanisms have been proposed to explain stars which do not seem to originate in the Galactic Centre \citep[e.g.][]{Yu2003,Przybilla2008, Herber2008, Tillich2009}. HVSs and HiVel stars can be used to infer many aspects about the Milky Way such as Galactic escape speed, and Galactic mass \citep{Smith2007,Piffl2013}, and HVS, in particular, offer a window into the mass function and dynamics of the environment near the massive black hole at the Galactic center \citep{PortegiesZwart2006, Lockmann2008a, Lockmann2008, Lu2007, Brown2012}. The benefit of finding and characterizing these HiVel stars and HVS can be translated into better understanding the structure, dynamics and evolution of the Milky Way. Most of the confirmed HVSs are early-type O and B type stars due to the selection bias of current HVS surveys. In recent years, there have been a few HVS candidates that are more evolved, later-type stars \citep{Brown2012, Brown2014, Palladino2014}. RAVE mostly targets late-type dwarfs and giants, and thus any HVS candidates will add to a now growing list of late-type HVSs candidates. Many studies have been devoted solely to search for HVS through dedicated surveys \citep[e.g.][]{Brown2012, Brown2014} as well as large surveys such as the Sloan Digital Sky Survey \citep[SDSS]{Kollmeier2010, Palladino2014, Zhong2014}. However, HVS have not been searched for in the RAVE data set. In this paper, we can naturally explore this by searching for HiVel stars.

While the main purpose of our study is to characterize, kinematically and chemically, HiVel stars, we will also investigate the runaway and hypervelocity candidates, for which we also have high-resolution spectra. In this paper, we start by searching for these rare HiVel stars using the RAVE data set and a series of selection criteria (described in section \ref{sec:data}). We then move on to discuss the kinematic, chemical distribution of our HiVel sample in section \ref{sec:results}. Using these distributions, we search for and suggest the origins of HVS and runaway star candidates. We summarize our key findings, put our findings into context with other studies and conclude in section \ref{sec:conclusion}. 
\section{A Sample of High-Velocity Stars}
\label{sec:data}
\subsection{RAVE Survey Data Release 4}
\label{subsec:RAVEDR4}
One of the easiest ways to search for HiVels is to use large astronomical surveys with high quality kinematic measurements, such as radial velocity (RV) and proper motions, etc. To this end, we make use of the fourth data release of RAVE \citep[RAVE DR4]{Kordopatis2013}. RAVE has obtained data using the multi-object spectrograph on the 1.2-m Australian Astronomical Observatory's UK Schmidt Telescope in Australia. The RAVE DR4 catalogue has reduced spectra and radial velocities (RVs) for nearly a half-million stars and represents one of the largest available catalogues to date. RAVE spectra are moderate resolution (R = $ \lambda/\Delta\lambda \sim$ 7000) around the Ca II triplet covering the wavelength range of 8410 -- 8795 \AA. For more information regarding RAVE, we refer the reader to the data release papers: \cite{Steinmetz2006, Zwitter2008, Siebert2011, Kordopatis2013} and the review of \cite{Kordopatis2014}. In addition to the accurate RV estimates, with errors on the order of 2 km~s$^{-1}$, DR4 contains stellar parameters and distances with errors of about 10-20 percent. RAVE also has an associated chemical abundance catalogue, which provides estimates of 6 elements including Fe and several $\alpha$-elements \citep[described in][]{Boeche2011}. The chemical pipeline used a hybrid approach of inferring abundance using the curves-of-growth for different elements as well as a penalizing $\chi^{2}$ technique of synthetic grid matching. \cite{Boeche2011} was able to determine the abundances of \afe and iron with a mean error of $\sim \pm$ 0.2 dex. This work will refer to [M/H] as metallicity and the traditional [Fe/H] as the iron abundance. [M/H] $\sim$ [Fe/H] under the assumption that the star follows the standard $\alpha$-enrichment scheme observationally seen in the Milky Way:

\begin{enumerate}
\item $\left[ \alpha/Fe \right ] $ = 0.0 dex for [Fe/H] $\geq$ 0 dex
\item $\left[ \alpha/Fe \right ] $= -0.4 $\times$ [Fe/H] dex for -1 $\leq$ [Fe/H] $<$ 0 dex
\item $\left[ \alpha/Fe \right ] $ = 0.4 dex for [Fe/H] $<$ -1.0 dex
\end{enumerate}
If these criteria do not hold, the RAVE metallicity does not equal [Fe/H]. The trend for RAVE stars is such that [M/H] $>$ [Fe/H] for low metallicity \cite[Figure 28]{Kordopatis2013}.

\subsection{High-Resolution Data}
\label{subsec:highres}
A few targets, namely several interesting HiVel and HVS candidates were followed up in high-resolution to enable detailed elemental abundance analysis. The high-resolution (R $\sim$ 31500) spectra were obtained using the ARC Echelle Spectrograph (ARCES) on the Apache Point Observatory (APO) 3.5-m telescope. The spectra were reduced in the standard way: bias subtraction, extraction, flat field division and stacking using the echelle package of IRAF\footnote{Distributed by NOAO, operated by AURA under cooperative agreement with the NSF.}. The final high-resolution spectra have a typical SNR $\sim$ 90-200 pixel$^{-1}$ in the wavelength region of 4500 -- 9000~\AA. 

Stellar parameters (T$_{\mathrm{eff}}$, log g, microturbulent velocity, $\xi$, and [Fe/H]) have been derived spectroscopically using the Brussels Automatic Code for Characterizing High accUracy Spectra (BACCHUS, Masseron et al., in preparation) code. The current version uses a grid of MARCS model atmospheres \citep{Gustafsson2008}, a specific procedure for interpolating the model atmosphere thermodynamical structure within the grid \citep{Masseron2006} and the radiative transfer code TURBOSPECTRUM \citep{Alvarez1998, Plez2012}. Atomic lines are sourced from VALD, \cite{Kupka1999}, \cite{Hill2002}, and \cite{Masseron2006}. Linelists for the molecular species are provided for CH (T. Masseron et al. 2014, in press), and CN, NH, OH, MgH and C2 (T. Masseron, in prep); the lines of SiH molecules are adopted from the Kurucz linelists and those from TiO, ZrO, FeH, CaH from B. Plez  (private communication). 

The stellar parameters determination relies on a list of selected Fe lines. The first step consists in determining accurate abundances for the selected lines using the abundance module for a given set of \Teff and log g. The abundance determination module proceeds in the following way:  (i) a spectrum synthesis, using the full set of (atomic and molecular) lines, is used for local continuum level finding; (ii) cosmic and telluric rejections are performed; (iii) local signal-to-noise is estimated; (iv) a series of flux points contributing to a given absorption line is selected. Abundances are then derived by comparing the observed spectrum with a set of convolved synthetic spectra characterized by different abundances.  Four different diagnostics are used: $\chi^2$ fitting, core line intensity comparison, global goodness-of-fit estimate, and equivalent width comparison. A decision tree then rejects the line, or accepts it keeping the best matching abundance. 

The second step consists in deducing the equivalent widths of Fe lines using the stellar parameter module. The last step of the procedure consists in injecting the derived equivalent widths in TURBOSPECTRUM to derive abundances for a grid of 27 neighboring model atmospheres (including three \Teff, three log g and three microturbulence velocities, covering the parameter space of interest). For each model, the slopes of abundances against excitation potential and against equivalent widths, as well as Fe I and Fe II lines abundances are computed. The final parameters are determined by forcing that the ionization equilibrium is fulfilled, and that simultaneously null slopes for abundances against excitation potential and against equivalent widths are encountered. The whole procedure is iterated twice per star, a first guess using the RAVE stellar parameters as a starting point and then again with a different starting point. This was done to obtain an independent set of stellar parameters compared to RAVE. We adopted the parameters from the second iteration for each star, however both iterations produce parameters that are in very good agreement. Individual elemental abundances in each of the absorption features were determined using a $\chi^{2}$ minimization technique to synthesized spectra. We visually inspected all fits in order to ensure the abundances were determined accurately. We take the mean and dispersion of the individual line abundances as the abundance and internal error respectively.

\subsection{Distances and Proper Motions}
\label{subsec:distancepm}
Proper motions were sourced from the fourth US Naval Observatory CCD Astrograph Catalogue (UCAC4), which contains proper motions for over 100 million objects. We choose UCAC4 because the error in proper motion is generally smaller compared to other catalogues \cite[see][for discussion]{Binney2014}. UCAC4 reaches a limiting magnitude of R = 16, with a peak in the formal uncertainty on the order of 4 mas/year \citep{Zacharias2013}. The small uncertainties in proper motion make them ideal for estimating the total velocity vector accurately. Following the suggestion of \cite{Zacharias2013}, we discard any star with an uncertainty in proper motion larger than 10 mas/year to avoid stars which have been labeled as `problematic' by the UCAC4. For most stars the UCAC4 proper motions are, within the errors of proper motion, in good agreement with other proper motion catalogues. Stars with no proper motions were discarded because the full kinematics cannot be explored without estimated proper motions. All stars that had a double star flag not equal to zero (it was either identified as a component of a double star system or it could not be ruled out as a double star) were excluded. Distances were determined from the estimated spectrophotometric parallax for each star using a method described by \cite{Binney2013}. We selected only stars that have estimated parallaxes because we need the distance to study the full kinematics of the stars. We expect typical uncertainties in parallax for our sample are on the order of 25 per cent \citep[see][for more details]{Binney2013}.

\subsection{Selection of High-Velocity Stars}
\label{subsec:selection}
To obtain robust data, we employed the quality control cuts laid out by \cite{Kordopatis2013} as well as some additional cuts described here.
\begin{enumerate}
\item The signal-to-noise ratio (SNR) must be larger than 20~pixel$^{-1}$. This cut was chosen to ensure that we are selecting out spectra that have well known uncertainties in the stellar parameters, and chemical abundances. Where there is more than one entry in the database we accept the entry with the highest SNR. 
\item The errors in the heliocentric RV (HRV) must be lower than 10~km~s$^{-1}$. This is to obtain precise RV measurements in order to constrain the full space motion. This cut is also necessary for accurate parameter estimation \citep{Kordopatis2013}. 
\item The estimated log g must be larger than 0.5~dex. Stars with log g less than 0.5~dex are much more likely to be treated in a problematic way during the stellar parameter estimation \citep{Kordopatis2013}. 
\item Calibrated metallicity [M/H] must be larger than -5 dex (measured by the stellar parameter pipeline). Stars below this were discarded stars because the stellar parameter estimation \citep{Kordopatis2013} pipeline was not designed to perform for these stars.
\item Calibrated \Teff must be between 4000 K and 7000 K. This was based on the range bounded by the synthetic library with which the RAVE chemical pipeline used.
\item The estimated stellar rotation velocity of the star, V$_{\mathrm{rot}}~<~ 50$~km~s$^{-1}$. This cut allowed us to discard stars that the stellar parameter and chemical pipelines would be likely to not produce reliable results.
\item All spectroscopic morphological flags defined by \cite{Matijevic2012} must be `normal'. This criterion minimized the chance of binary star contamination or highly uncertain stellar parameters and distances \citep{Kordopatis2013, Binney2013, Boeche2011}. 
\item The DR4 algorithm convergence parameter (algo\_conv) must not equal 1. We used this cut to ensure the stellar parameter pipeline converged \citep{Kordopatis2013}
\item The value frac (i.e. the fraction of the spectrum that matches the model in a satisfactory way) associated with the chemical pipeline must be greater than 0.7.
\item  The $\chi^2$ associated to the chemical pipeline must be lower than 2000. This cut along with the value frac cut was used to confirm that the spectra were adequately fit with no glaring errors in the continuum \citep{Boeche2011}.
\item The Tonry-Davis correlation coefficient, which is a measure of the quality of the template fit for each stellar spectra during the RV measurement, must be larger than 10  \citep{Piffl2013, Steinmetz2006, Kordopatis2013}. 
\end{enumerate}
 A total of 274 481 objects passed the 11 quality cuts described above. We corrected the HRVs for solar and local standard of rest (LSR) motion to obtain a RV relative to the Galactic rest-frame (GRV) using the following formula:
\begin{equation}
\mathrm{GRV} = \mathrm{HRV} + (U_{\sun} \cos{l} + (V_{\sun} + V_{\mathrm{LSR}}) \sin{l}) \cos{b} + W_{\sun} \sin{b},
\end{equation}
 
where $U_{\sun}, V_{\sun}, W_{\sun}, V_{\mathrm{LSR}}, l, b$ are the 3-dimensional solar velocity, the velocity of the local standard of rest (assumed to be 220 km~s$^{-1}$), the Galactic longitude and latitude, respectively. For reference, the velocity convention adopted by this work is: U is positive if directed toward the Galactic Centre ($l$ = 0$^{\circ}$, $b$ = 0$^{\circ}$), V is positive along the direction of rotation ($l$ = 90$^{\circ}$, $b$ = 0$^{\circ}$) and W is positive if pointed toward the North Galactic Pole ($b$ = 90$^{\circ}$). In this convention, the Sun's orbital velocity vector $\vec{v}_{\sun} = [U_{\sun}, V_{\sun}, W_{\sun}] = [14.0, 12.24, 7.25]$ km~s$^{-1}$, V$_{\mathrm{LSR}}$ = 220 \kms and position = [8.28, 0, 0] kpc  \citep{Schonrich2012}. 

 \begin{figure}
  \includegraphics[scale=0.3]{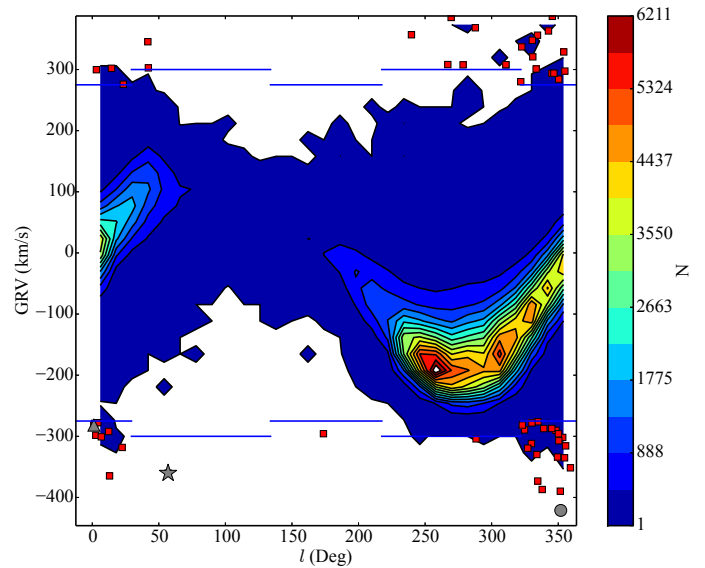}
  \
 \caption{The selected HiVel targets (red squares) in $l$-Galactic relative radial velocity (GRV) space. The blue lines indicate the selection criteria where the |GRV| $>$ 300 km s$^{-1}$ near galactic latitudes pointed in the direction of disk rotation ($l = 90  \pm 50  ^{\circ} $ and $270  \pm 50 ^{\circ}$) and |GRV| $>$ 275 km s$^{-1}$ elsewhere. The contours show the full RAVE DR4 sample from which the HiVel targets were selected. The color in all 2D density diagrams, like this one, represents the number of stars in each density contour for the full RAVE dataset. The gray circle, star and triangle refers to three interesting targets, namely J1544, J2217, and J1610 respectively, that we discuss in the further detail in later sections.}
  \label{fig:lvsgrv}
  \end{figure}

We initially selected objects with an absolute |GRV| $>$ 300 km~s$^{-1}$ in regions where most of the disk velocity is along the line-of-sight (at $l$ = 90 $\pm 50 ^{\circ} $ and 270 $\pm 50 ^{\circ}$ ) and |GRV| $>$ 275 km~s$^{-1}$ elsewhere. This requirement was designed to cut out most of the disk contamination that would otherwise occur owing to the geometry. These boundaries are selected to be just above the speed one would expect when considering disk rotation and velocity dispersion. Lowering the cut on GRV would cause larger contamination from ordinary disk stars while increasing the threshold decreases the total sample size. The selection of our sample can be found in Figure \ref{fig:lvsgrv}. The open circle, star and triangle refers to three interesting targets including J154401.1-162451, J221759.1-051149, and J161055.6-112009 (henceforth J1544, J2217, and J1610) respectively, that we discuss in the further detail in later sections. The signal of the disk can be seen as the sinusoidal-like high-density path seen in Figure \ref{fig:lvsgrv}. This is a geometric effect caused by the fact that disk stars move along rotation and thus when we observe at angles directed towards or away from rotation, most of the disk star velocity will be in the line-of-sight direction. The high density around $l = 270 ^{\circ}$ with GRV $\sim$ --220~km~s$^{-1}$ represents the disk because at that Galactic longitude, the total velocity vector is primarily in the line-of-sight direction.  

\begin{figure}
 \includegraphics[scale=0.28]{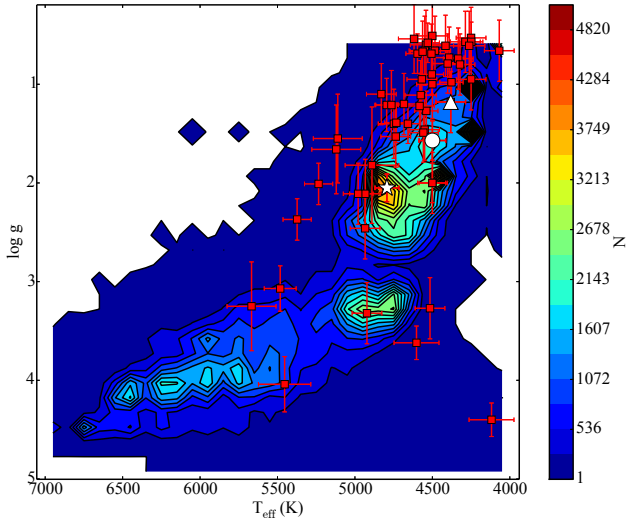}
 \caption{Contours of the \Teff - log g for stars in the RAVE sample with the HiVel stars (red squares) shown as being primarily giant stars. The open circle, star and triangle are the same as in Figure \ref{fig:lvsgrv}.}
  \label{fig:Tefflogg}
  \end{figure}

\begin{figure}
\includegraphics[scale=0.28]{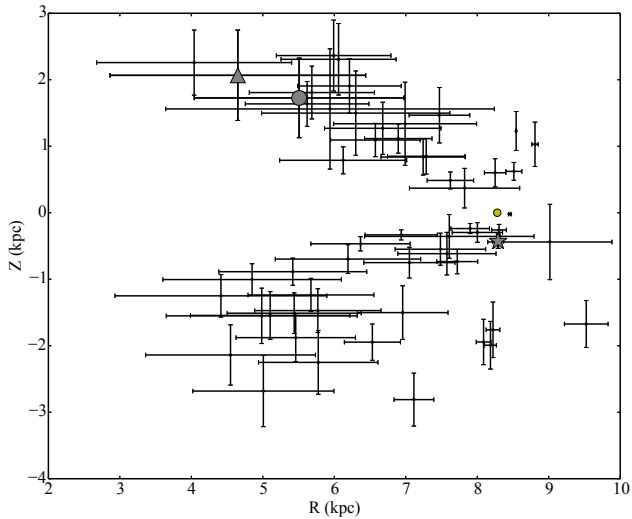}
 \caption{The position of our HiVel stars relative to the Galaxy. For reference the Sun (yellow circle) is at (R, Z) = (8.28, 0) kpc. The gray circle, star and triangle are the same as in Figure \ref{fig:lvsgrv}.}
  \label{fig:RvsZ}
  
  \end{figure}

We further required the distances, proper motions (with uncertainties less than 5 mas/yr) and metallicities to be known. There are 57 stars in our final sample that passed all of the 11 quality control cuts and the kinematic cut described above. \cite{Piffl2013} used a sample of 76 HiVel stars in RAVE to determine the mass of the Milky Way halo. Our sample has fewer HiVel stars than \cite{Piffl2013} which is a result of the fact that they use a much lower GRV cut (200 km~s$^{-1}$) than we do by selecting for a counter-rotating population assumed to be the halo population. Our kinematic selection criteria in $l$-GRV space can be seen in Figure \ref{fig:lvsgrv}. It is important to note that we choose to select our HiVel stars on the GRV rather than the full space motion because of the smaller error in GRV, on the order of a few km~s$^{-1}$, compared to the error on the full space motion (few tens of km~s$^{-1}$). Most of the HiVel stars in our sample are giant stars (Figure \ref{fig:Tefflogg}) and located within 5 kpc of the Sun (Figure \ref{fig:RvsZ}). RAVE is run out of the Southern Hemisphere and as a result the spatial distribution of the selected HiVel stars is not symmetric around the Sun. 

\subsection{Full Space Velocity and Stellar Orbits}
\label{subsec:orbits}
We fully resolve the space velocity and position vectors in order to study the orbital parameters of our sample. The right ascension ($\alpha$), declination ($\delta$), and distances for each source yield a full 3D position vector relative to the Galaxy. These vectors are related by the Sun's Galactocentric position ($\vec{r}_{\odot} = [8.28, 0, 0]$ kpc) such that $\vec{r} = \vec{r}_{*} + \vec{r}_{\odot}$ \citep{Schonrich2012}. Furthermore, the proper motions give two dimensions of velocity within the plane of the sky, $\vec{\mu}$. We use the estimated proper motion, distance, and RV to construct a current velocity vector (relative to the Galactic-rest frame) via \cite{Johnson1987}. Uncertainties in the current position and velocity vector were determined using a Monte Carlo approach with a thousand realizations, randomly varying the uncertainties of all of the observables and recalculating the position and velocity vectors and studying the final distribution, similar to the approach used by \cite{Gratton2003} and \cite{Boeche2013}. We excluded two stars with uncertainties in the total space velocity larger than 200 km s$^{-1}$ as it would be difficult to constrain their total space motions. 

To obtain the orbital parameters for our HiVel sample, we integrated the orbit of a test particle through an assumed Galactic potential ($\Phi$) which is a sum of the potential of a logarithmic halo ($\Phi_{halo}(r)$), Miyamoto-Nagai disk ($\Phi_{disk}(R,z)$), and a Hernquist bulge ($\Phi_{Bulge}(r)$). We made use of the same parameter choices as \cite{Johnston1995}. 
\begin{equation}
\Phi_{halo}(r) =\frac{v_{0}^{2}}{2}\ln{(r^{2} + d^{2})} ,
\end{equation} 
where $v_{0}$ is a characteristic velocity of 186 km~s$^{-1}$ with a scale length, $d$, of 12.0 kpc.   
\begin{equation}
\Phi_{disk}(R,z) = -\frac{GM_{disk}}{\sqrt{R^2 + \left (a+\sqrt{z^2+b^2}  \right )^2}} ,
\end{equation} 
where the $M_{disk}$ is the mass of the disk assumed to be 10$^{11}$ M$_{\sun}$), $a$ and $b$ are scale lengths set to 6.5 kpc, and 0.26 kpc, respectively. 
\begin{equation}
\Phi_{Bulge}(r) = -\frac{GM_{Bulge}}{r+c} ,
\end{equation} 
where $M_{Bulge}$ is the mass of the bulge and is set to $3.4 \times 10^{10} M_{\sun}$ and $c$ is a scale-length set to 0.7 kpc. In the above definitions $r$ = $\sqrt{x^2+y^2+z^2}$ and $R$ = $\sqrt{x^2+y^2}$. Using this potential, we confirmed the circular speed, $v_{\mathrm{circ}}$, at the solar radius of 8.28 kpc to be $v_{\mathrm{circ}}$ = 224 \kms and an orbital period for the local standard of rest (LSR) at this radius of 220 Myr consistent with \cite{Schonrich2012}. We also verified that energy and angular momentum is conserved in all orbital integrations to at least one part in a million or better. 

To better understand the kinematics of our stars, we estimated the maximum distance above the Galactic plane (denoted Z$_{max}$) and the eccentricity from the orbital integration. We define the eccentricity as a function of the apogalactic distance, r$_{ap}$, and the perigalactic distance, r$_{per}$, such that $e$ = (r$_{ap}-$r$_{per}$)/(r$_{ap}+$r$_{per}$). Uncertainties in the orbital integrations were estimated in a similar Monte Carlo approach as above (the initial conditions were varied to within their uncertainties) with 100 orbital integrations. We find the uncertainty in the eccentricity is less than 0.15. 

\section{Results: Metal-Poor High-Velocity Stars, Ejected Disk Stars and Hypervelocity Stars}
\label{sec:results}
In this section we discuss the kinematic (section \ref{subsec:kinematics}) and chemical (section \ref{subsec:chemistry}) distributions of our HiVel star sample. By combining the results of these two sections we discuss the discovery of a metal-rich halo star that likely originated in the Galactic disk and put forward a HVS candidate. 

\subsection{Kinematics of High-Velocity Stars}
\label{subsec:kinematics}
\begin{figure}
 \includegraphics[scale=0.28]{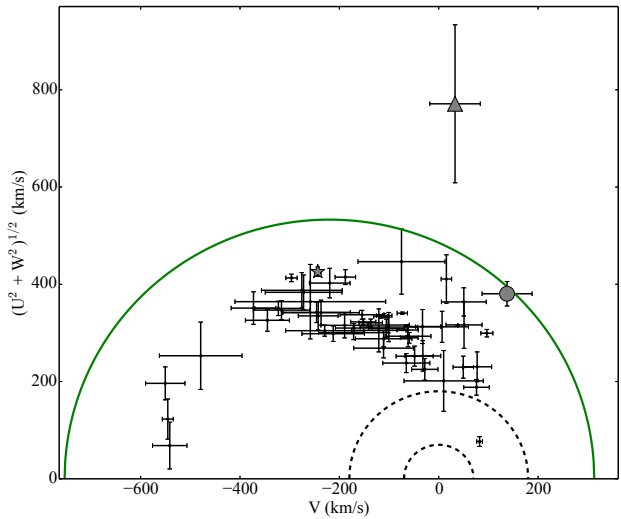}
 \caption{Toomre diagram for the HiVel sample. All velocities are relative to the LSR. The 2 black dashed rings show roughly the boundaries of the thin disk and thick disk at a constant velocity of 70~km~s$^{-1}$, 180~km~s$^{-1}$ respectively \citep{Venn2004}. We can see that most of our HiVel stars belong to the halo kinematically. The green solid line represents a constant galactic-rest frame velocity of 533~km~s$^{-1}$, and is thus shifted relative to the other velocities. A star above this boundary may be HVS candidate pending its position. We find one HVS candidate with a total Galactic rest frame velocity larger than 800 \kms (more than 1-$\sigma$ above the escape speed). There are two stars, namely J1610 and J154401.1-16245, which have velocities above HVS limit. The gray circle, star and triangle are the same as in Figure \ref{fig:lvsgrv}.}
  \label{fig:toomre}
  \end{figure}

We first studied the kinematics of our HiVel sample using a Toomre diagram (Figure \ref{fig:toomre}), which quantifies different Galactic components using the velocity vector. It is important to note that the velocities in the Toomre diagram are relative to the LSR and thus the HVS boundary (green line, Figure \ref{fig:toomre}), which must be converted to a non-rotating reference frame, was shifted by V = --220~km~s$^{-1}$.  From inspection, there are two stars which sit at or above the HVS boundary (green line, Figure \ref{fig:toomre}). We note that the HVS boundary is position dependent. For simplicity, we choose the escape speed at the solar circle as an illustrative HVS boundary on the Toomre diagram. However, a true HVS candidate must have a velocity higher than the escape speed at its position. There is also one star that has disk-like kinematics and is likely with the high-velocity tail of the thin disk or disk-like contaminants. Using the rough boundaries that kinematically separate the thin-disk, thick disk and halo of \cite{Venn2004} for example, we found that most of our HiVel stars exist in the halo-region of the Toomre diagram which means these stars are likely be on highly elliptical orbits reaching out to a maximum distances from the Galactic plane, Z$_{max}$, larger than 10 kpc with eccentricities of $e \gtrsim$ 0.5. Adopting larger boundaries in the Toomre diagram that kinematically separate the thin-disk, thick disk and halo would result in a slight contamination by the thick disk. 

\begin{figure}
 \includegraphics[scale=0.4]{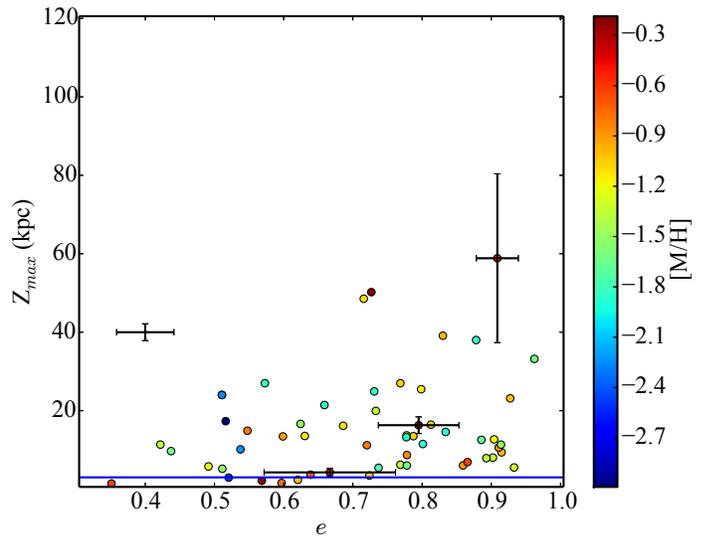}
 \caption{Eccentricity, $e$, as a function of the maximum Galactic plane height obtained by our HiVel stars during a 1 Gyr orbital integration. The color of each star represents it metallicity, [M/H]. The solid horizontal line represents the edge Z$_{max}$ of the thick disk, Z = 3 kpc \citep{Carollo2010}. The error bar to the left represents the median uncertainty in both parameters. The other three error bars in $e$ and Z$_{max}$ are shown for the stars with the largest errors on distances. This plot excludes HVS candidates. The high Z$_{max}$ and $e$ for most of the HiVel stars indicate they are consistent with the Galactic halo. }
 \label{fig:ezmax}
 \end{figure}

We further studied the kinematics of our sample by comparing the $e$ and Z$_{max}$ (Figure \ref{fig:ezmax}). The power of the $e$-Z$_{max}$ plane is the ability to sort out stars of similar orbits, because $e$ describes the shape of the orbit and Z$_{max}$ describes the amplitude of the vertical oscillations \citep[e.g.][]{Boeche2013}. The $e$-Z$_{max}$ plane combined with metallicity allows us to characterize the orbits of our HiVel stars while also considering the chemical distribution. Our HiVel sample has median $e$ of 0.73 and median Z$_{max}$ of 13 kpc, which is kinematically consistent with the halo population. This result confirms the assumptions of older works \citep[e.g.][]{Schuster1988, Ryan2003, Schuster2006} that HiVel stars in the solar vicinity mostly belong to the halo. We note that while our specified Galactic potential is thought to be an adequate assumption locally, the potential at large distances from the Sun can be relatively uncertain. Since HiVel stars can probe these distant regimes the uncertainties in the orbital parameters, namely the Z$_{max}$, r$_{ap}$ and r$_{per}$, are probably underestimated as we only quote the uncertainties by propagating the errors on the observables. There are a few stars with Z$_{max} \lesssim$ 3 kpc and eccentricities below 0.6. These stars could be interpreted as thick-disk contaminants especially given their relatively high ($>$~--0.90~dex) metallicities \citep{Boeche2013, Kordopatis2013b, Kordopatis2013AA}. 

\subsection{Chemical Distribution of High-Velocity Stars}
\label{subsec:chemistry}
\begin{figure}
 \includegraphics[scale=0.33]{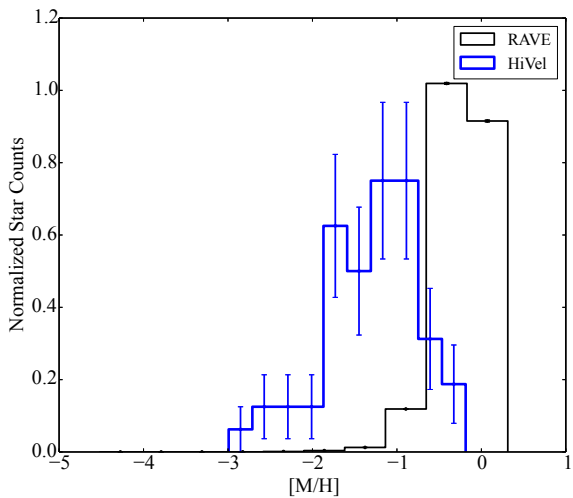}
 \caption{[M/H] distribution for the RAVE catalogue (black) and the HiVel sample (blue). The HiVel stars in our sample are significantly more metal-poor compared to the RAVE mother sample. The peak of the metallicity distribution of HiVel stars is --1.18 dex and is consistent with the Galactic halo. The error bars are computed assuming Poisson noise. }
  \label{fig:metdist}
  \end{figure}

The kinematics of our HiVel sample (Section \ref{subsec:kinematics}) indicate these stars are drawn from the halo population and thus they should also have a chemical fingerprint that is consistent with the halo. In Figure \ref{fig:metdist} we compare the normalized metallicity distribution of the RAVE and HiVel samples. It is clear that the mean metallicity of the RAVE sample ($\overline{[M/H]}_{\mathrm{RAVE}}$ = --0.22 dex) is significantly higher than mean metallicity of the HiVel sample ($\overline{[M/H]}_{\mathrm{HiVel}}$ $\sim$ --1.2 dex). The mean metallicity of our HiVel sample is slightly higher but consistent within the errors with the inner Galactic halo, which is thought to have a mean metallicity around --1.60 dex \citep{Carollo2007, Carollo2010}. The inner Galactic halo is also thought to have measurable $\alpha$-enhancement \citep{Nissen2010, Haywood2013, Boeche2013}. 

The distribution in \afe space will provide information on the birthplace of the stars. Many recent surveys have shown that the different components of the Galaxy can be partially separated in [$\alpha$/Fe] - metallicity space \citep{Nissen1997, Fulbright2002, Stephens2002,Nissen2010, Ruchti2010, Nissen2012, Haywood2013, Feltzing2013}. These studies show a relationship between the metallicity and [$\alpha$/Fe] that is described in section \ref{subsec:RAVEDR4}.

\begin{figure}
  \includegraphics[scale=0.28]{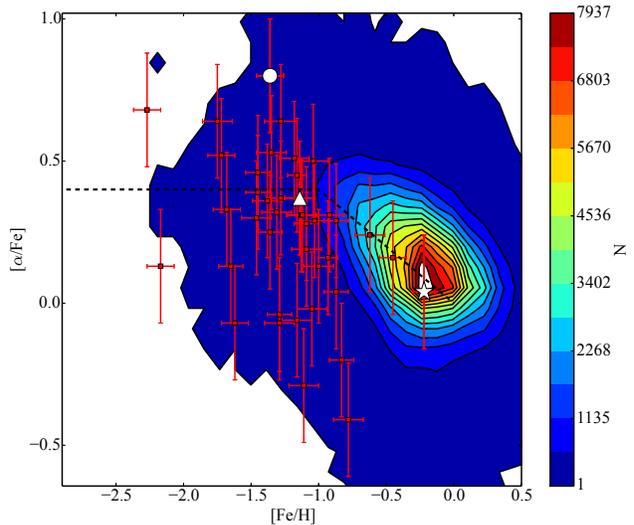}
 \caption{Contour plot showing the [$\alpha$/Fe] - [Fe/H] for RAVE including where the HiVel stars (red squares) fall in this space. The dotted black line represents the standard Galactic trend in this space. Most of our HiVel stars are, within the errors consistent with the halo population with [Fe/H] less than -1.0 dex and noticeable $\alpha$-enrichment. The open circle, star and triangle are the same as in Figure \ref{fig:lvsgrv}.}
  \label{fig:alphamet}
  \end{figure}

\begin{figure}
   \includegraphics[scale=0.33]{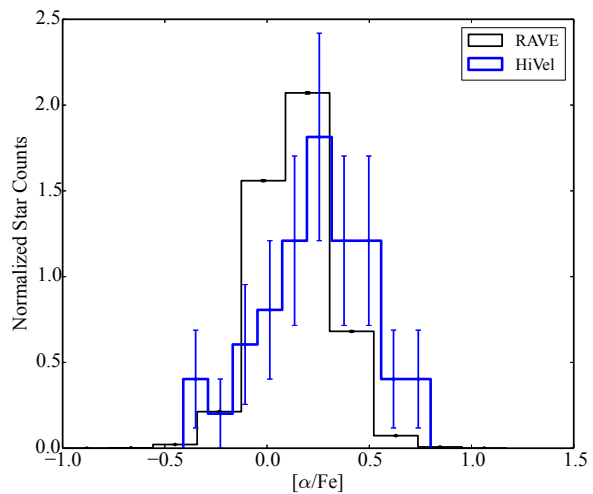}
 \caption{[$\alpha$/Fe] distribution for the RAVE catalogue (black) and the HiVel sample (blue). The HiVel sample is slightly more $\alpha$-enriched compared to the RAVE mother sample but the dispersions are comparable.}
  \label{fig:alphadist}
  \end{figure}

For comparison, we plot our HiVel stars in this space relative to the full RAVE sample along with the expected Galactic trend (see Figure \ref{fig:alphamet}). The expected uncertainty in both [M/H] and [$\alpha$/Fe] is  $\sim \pm$ 0.2 dex \citep{Kordopatis2013}.  Further, we can see from Figure \ref{fig:alphadist} that our HiVel star sample is slightly $\alpha$-enriched, with a mean $\alpha$-abundance of $\overline{ [\alpha/Fe]}_{HiVel}$ = +0.24 dex, compared to the RAVE mother sample, with a mean $\alpha$-abundance of $\overline{ [\alpha/Fe]}_{RAVE}$ = +0.14 dex. This result is, within the errors, chemically consistent with the halo population. The large dispersion (on the order of 0.25 dex) in the [$\alpha$/Fe] is likely a result of the uncertainty of the individual [$\alpha$/Fe] estimates, but may also represent an $\alpha$-poor and $\alpha$-enriched population in our HiVel sample. The large uncertainty in the [$\alpha$/Fe] estimates, particularly at low metallicity, is a result of insufficient spectral information on the abundance of $\alpha$-elements in the part of the spectrum covered by RAVE \citep{Kordopatis2013}. 

 \begin{figure}
 \includegraphics[scale=0.28]{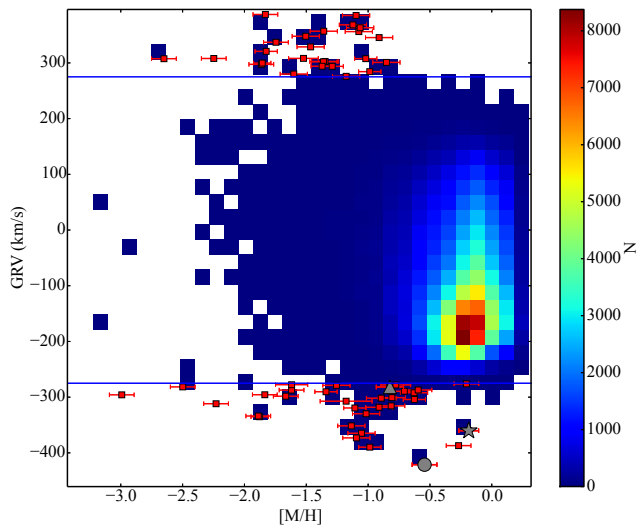}
 \caption{2D density plot of the GRV and [M/H] for the RAVE sample. The horizontal lines indicate the adopted kinematic minimum kinematic GRV needed to be classified as a HiVel candidate (i.e. absolute GRV $>$ 275~km~s$^{-1}$). It is interesting to note that there are three HiVel stars with relatively high metallicities ([M/H] larger than -0.5 dex). The gray circle, star and triangle are the same as in Figure \ref{fig:lvsgrv}.}
  \label{fig:grvmet}
  \end{figure}

Combining the kinematic and chemical properties, we plot the 2-dimensional density of the GRV as a function of metallicity for the RAVE sample and our HiVel stars (red squares) in Figure \ref{fig:grvmet}. Viewing the results in this space allows us to identify clearly one star, J2217 that has an extremely high GRV ($\sim$ --360~km~s$^{-1}$), but paradoxically is metal-rich ([M/H] = -- 0.18 $\pm$ 0.08 dex). This star is an outlier compared to the rest of the metal-poor HiVel stars of our sample. If we make a simplistic assumption that the (inner) Galactic stellar halo metallicity distribution function can be modeled as a Gaussian with a mean of [M/H] = --1.50 dex and $\sigma_{[M/H]}$ = 0.50 \citep{Chiba2000}, the probability of drawing a star of that metallicity is 0.4 per cent (2.64$\sigma$). Assuming higher mean and dispersion values, \cite[-1.20, 0.54, see][]{Kordopatis2013b} the probability of drawing a star of that metallicity is 2.9 per cent (1.89$\sigma$). In either case, the probability of drawing a star of this metallicity from the Galactic halo population is small \citep[$<$ 4 per cent][]{Carollo2007,Carollo2010, Kordopatis2013b, An2013}. This star provides us with a unique opportunity to explore metal-rich halo stars. As such, the next section is devoted to exploring this object in more detail.  

It is worth mentioning there are two additional stars that have [M/H] larger than --0.50~dex which are classified as HiVel stars. J152905.9-365544 is a giant star which has a GRV~= --~276 $\pm$ 1.5~\kms with a metallicity of [M/H] = --0.21~$\pm$~0.1 dex and [$\alpha$/Fe] $\sim$~+0.16~$\pm$~0.2~dex. On the Toomre diagram (Figure \ref{fig:toomre}), this star sits just above the thick disk region. An orbital integration of this star was performed and showed that this star has an $e \sim$ 0.5 with Z$_{max}$ $\sim$ 3 $\pm$ 1 kpc. Using the same probabilistic kinematic classification from \cite{Bensby2003} this star would be categorized as a thick disk star. Given the kinematic and chemical properties of this star, we expect it is a thick disk contaminate. The second star, J193647.0-590741, has an estimated metallicity of [M/H] = --0.27 $\pm$ 0.1~dex, yet kinematically it has a GRV = --387 $\pm$ 1.7~km~s$^{-1}$. Orbital integration for this star indicates it has an $e \sim$ 0.82 with Z$_{max} \sim 16 \pm$ 5 kpc which resembles halo-like properties. While the stellar parameter pipeline is able to estimate its \Teff, log g and [M/H], the chemical pipeline fails to provide an estimate of the [Fe/H] and [$\alpha$/Fe]. This could be due to the low SNR ($\sim$ 20) and thus high-resolution, high SNR follow-up will be necessary to confirm the chemical signature of this star. 

\subsection{Captured Star or High-Velocity Ejected Disk Star?: The Case of J2217}
The giant star J2217 represents an unusually fast-moving object with a metallicity significantly above --1.0 dex. The RAVE stellar parameter pipeline has estimated it to have a \Teff = 4790 $\pm$ 80~K and log g = 2.05 $\pm$ 0.15~dex. It has a metallicity of --0.18 $\pm$ 0.1~dex (at a SNR = 71.0) and an [$\alpha$/Fe] = +0.04 $\pm$~0.2 dex. The chemistry of this star, particularly the high metallicity and low levels of $\alpha$-enhancement, is consistent with a disk or possibly captured star but not the Galactic halo like most of the other HiVel stars. We have no reason to believe this star is a member of a binary star based on the `normal' spectral morphological classification of \cite{Matijevic2012} and is classified as a single normal star in the UCAC4 catalogue \citep{Zacharias2013}. 

  \begin{figure*}
  \includegraphics[width=2\columnwidth]{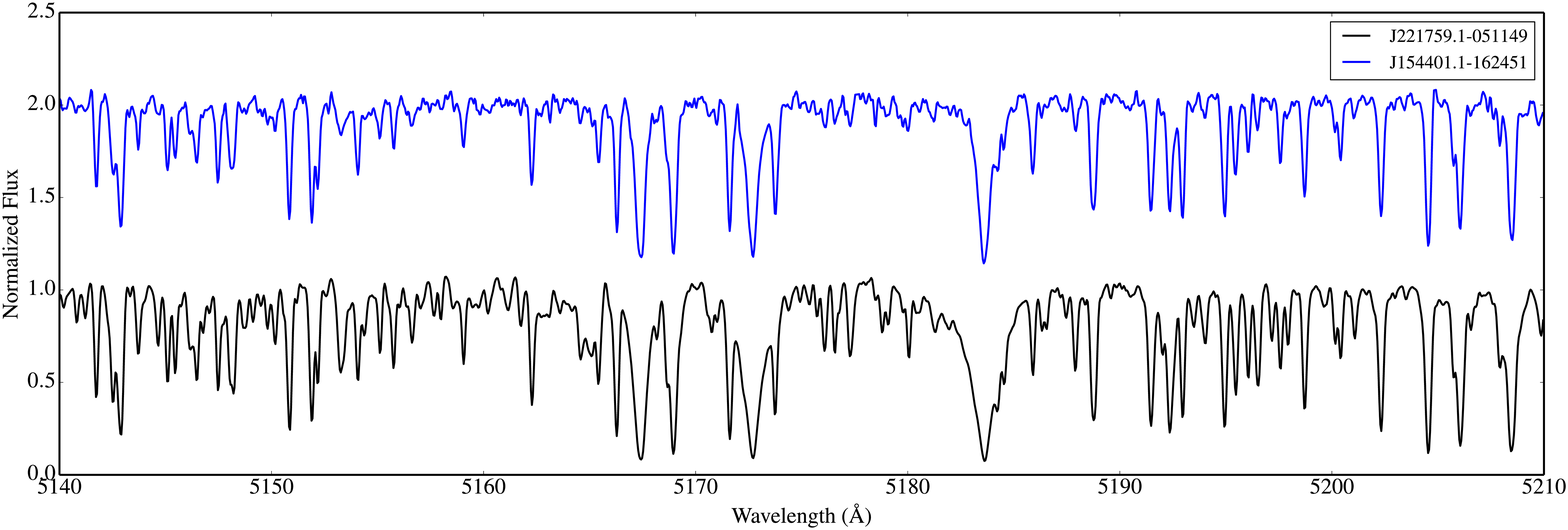}
 \caption{The observed high-resolution ACRES spectrum of the giant star J2217 (black line) and J1544 (blue line) in the Mg I triplet region. In both cases these spectra were used to determine the chemical abundances which are shown in Table \ref{tab:highreschemJ22} and Table \ref{tab:highreschem}.}
  \label{fig:higresspec}
  \end{figure*}
  
Kinematically, the star is in the extreme halo region residing below the HVS boundary on the Toomre Diagram (Figure \ref{fig:toomre}). The total Galactic rest frame velocity of this star is 426 $\pm$ 10 km~s$^{-1}$. The Tonry-Davis correlation coefficient estimated by RAVE is 65 indicating the template for cross-correlation was a good fit and the uncertainty in the measured RV approximately 1 km s$^{-1}$. This puts it $\sim$ 100 \kms below the Galactic escape speed of \cite{Piffl2013} at its Galactocentric distance of r = 8.01 $\pm$ 0.13 kpc. To better understand the kinematics of this object we integrated its orbit over 1 Gyr (Figure \ref{fig:orbitJ22175}) to get an idea of the shape of the orbit without being dominated by errors due to the observables. We found that J2217 reaches a Z$_{max}$ = 31 $\pm$ 5 kpc and has an overall eccentricity of $e$ = 0.72 $\pm$ 0.02 . This star kinematically resembles a halo star given its Z$_{max}$ \citep[e.g.][]{Coskunoglu2012} and eccentricity yet its chemistry suggests it may belong to the Galactic disk or a dwarf galaxy \citep[e.g.][]{Sheffield2012}.

The orbital integration described above was used to estimate the time-of-flight (TOF) for this star assuming it was produced in the Galactic disk and ejected into the halo (Figure \ref{fig:orbitJ22175}). The star is passing through the disk and the TOF required to get the star near the galactic disk again ($\left|z\right|<$ 3 kpc) is $\sim$ 600 Myr well within the lifetime of a low-mass star. \cite{Bromley2009} argued that the metal-rich tail of the (inner) halo metallicity distribution may come from stars born in the disk and kinematically heated (by binary supernova ejection) into the halo as ejected disk stars. In terms of the time-scale, kinematics, and chemistry it is perfectly reasonable this object was born in the Galactic disk and was kinematically heated, maybe as a result of binary ejection or tidal interactions with satellite galaxies as in \cite{Purcell2010} or other gravitational means, causing it to be now observed as a part of the Galactic halo. Alternatively, this star could be a captured star from a dwarf spheroidal galaxy. Reaching out to a vertical distance of 35 kpc is within the distance of a couple massive dwarf galaxies (e.g. Sagittarius dwarf galaxy). Chemically, massive dwarf galaxies like the Sagittarius dwarf spheroidal galaxy may contain some stars as metal-rich as [M/H] = -0.20 dex with a depletion in [$\alpha$/Fe] $\sim$ -0.20 dex \citep{Sbordone2007}. J2217 is metal-rich with no noticeable enhancement in [$\alpha$/Fe] (i.e. an [$\alpha$/Fe] = --0.20 dex is at the edge of the $\pm$ 0.2 dex error budget for the [$\alpha$/Fe]) estimate. However, given that the 3 dimensional velocity vector of this star indicates that it is currently on its way out of the Galaxy and chemically resembles the Galactic disk, it is more likely this star is an ejected disk star. 

\begin{figure*}     
       \subfigure[]{
               \includegraphics[width=\columnwidth]{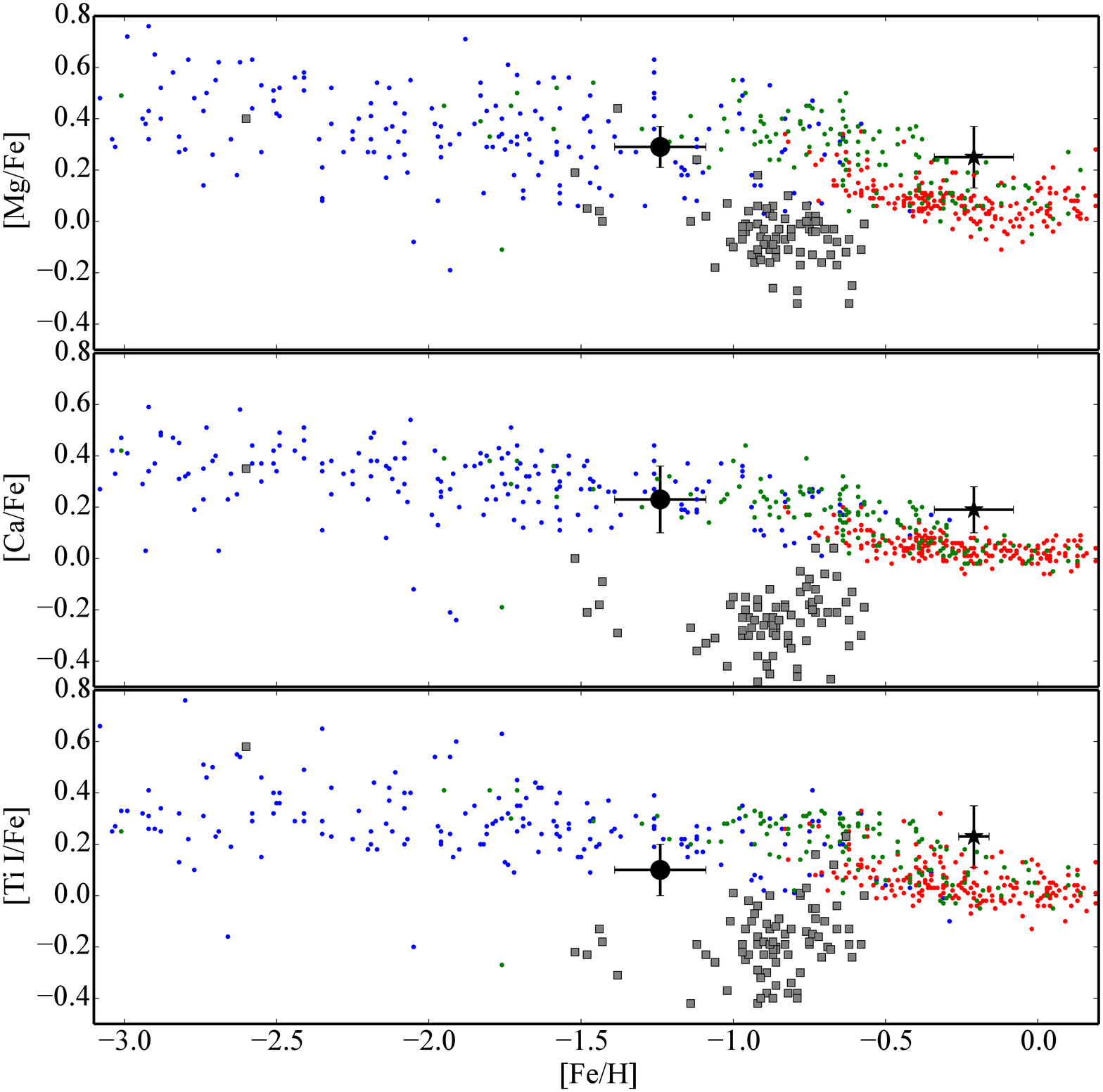}
                \label{fig:alphachem}  }
        \subfigure[]{
                \includegraphics[width=\columnwidth]{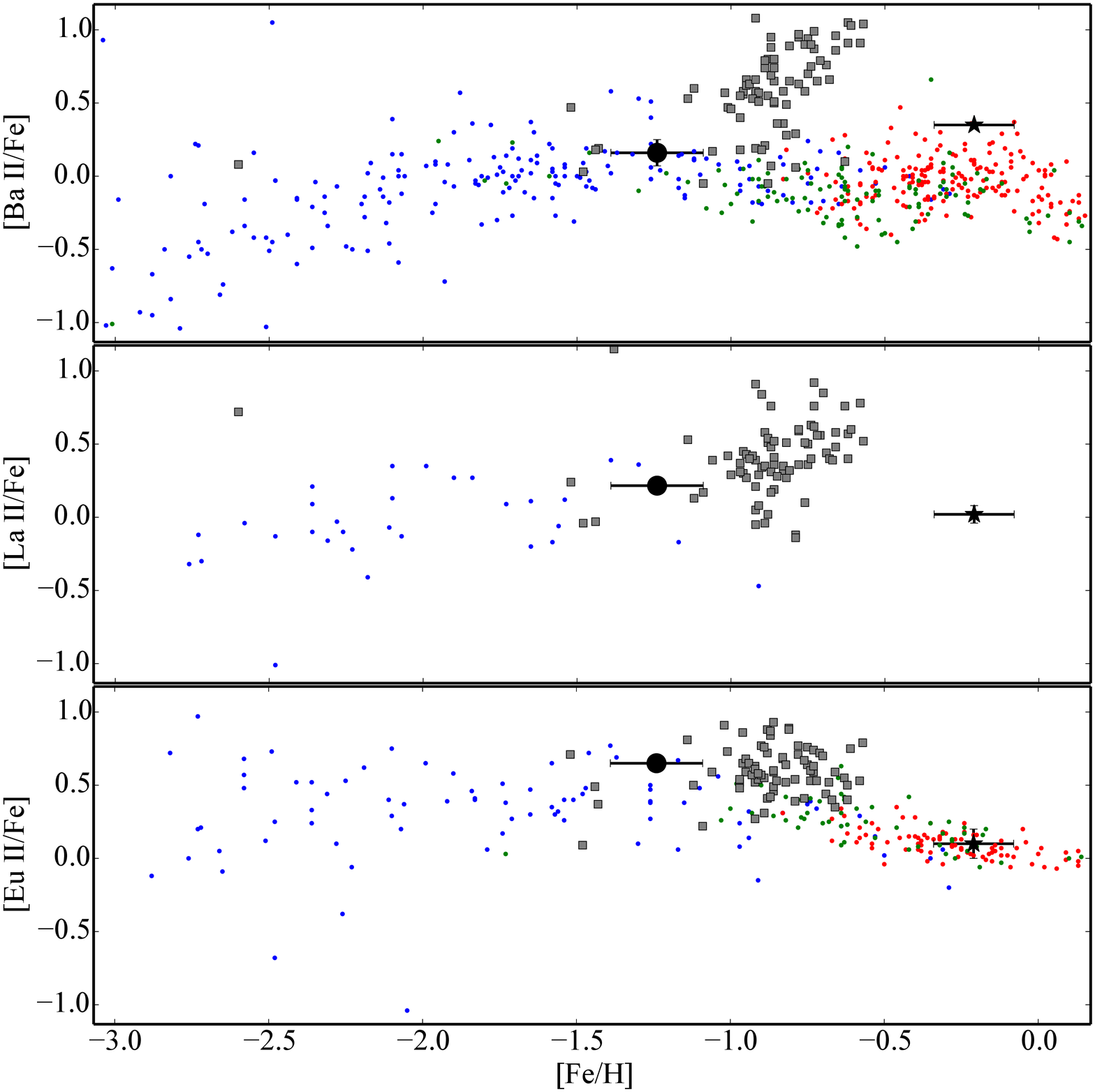}
               \label{fig:schem}   }

        \caption{(a) The abundance patterns of the $\alpha$-elements as a function of metallicity including [Mg/Fe], [Ca/Fe] and [Ti/Fe] from top to bottom, respectively for J2217 (black star) and J1544 (black circle). (b) The abundance patterns of neutron-capture elements as a function of metallicity including [Ba II/Fe], [La II/Fe], and [Eu/Fe] from top to bottom respectively. The small coloured circles represent abundances of thin disk (red), thick disk (green) and halo (blue) stars from \protect\cite{Venn2004}. For comparison, the grey squares show an example of a dwarf galaxy (Fornax) from \protect\cite{Letarte2010}. J2217 is chemically consistent in the $\alpha$-elements and most neutron-capture elements with the Galactic thick disk. On the other hand J1544 is chemically consistent with the halo field population or dwarf galaxies. The error bars on the side represent the mean error of the abundances from the literature.}
        \label{fig:alpha+sprocess}
\end{figure*}

\begin{figure}
  \includegraphics[width=\columnwidth]{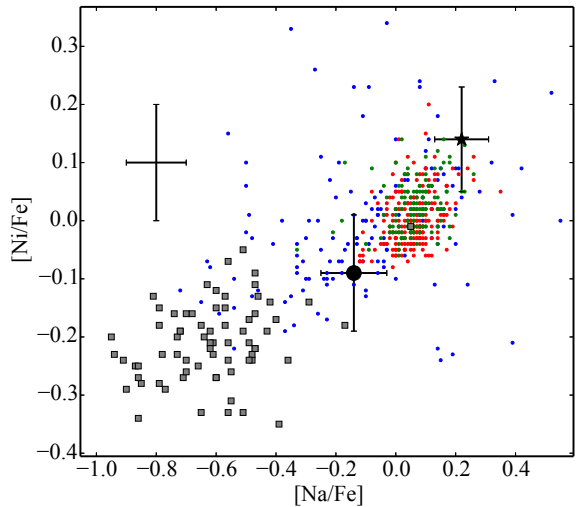}
 \caption{The observed high-resolution abundances of Ni as a function of Na for J2217 (black star) and J1544 (black circle). The symbols are the same as Figure \ref{fig:alpha+sprocess}. We find that J2217 is chemically consistent with the disk and not a dwarf galaxy (such as the Fornax). However it is possible that J1544, within the errors, may be consistent with a massive dwarf galaxy.  The error bars on the side represent the mean error of the abundances from the literature.}
  \label{fig:NaNi}
  \end{figure}

\begin{figure}
  \includegraphics[scale=0.3]{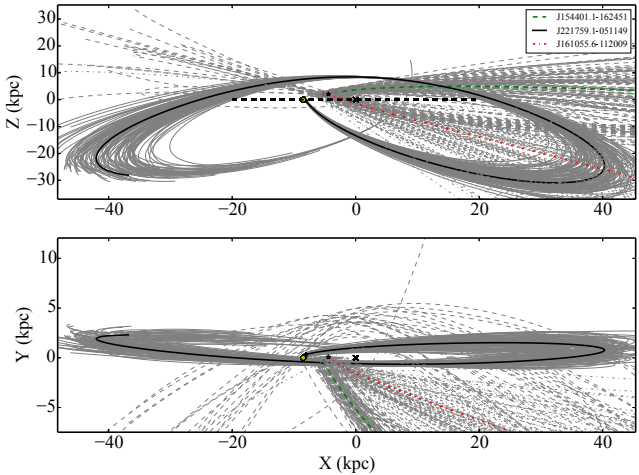}
 \caption{A 1 Gyr orbital integration for J2217 (black solid line),  J1544 (green dotted line), and J1610 (red dash-dotted line). The yellow circle represents the Sun, the black `x' represents the Galactic center and the black asterisk represents the current position of J2217 and J1544. The thin grey lines are 100 draws of the orbital integration to illustrate the uncertainty. We find that both stars are currently passing through the disk.}
  \label{fig:orbitJ22175}
  \end{figure}

We observed J2217 in high-resolution (black line in Figure \ref{fig:higresspec}) using the APO to try determining whether the star is captured or ejected. The analysis of the high-resolution spectra (described in Section \ref{subsec:highres}) yielded stellar parameters  (\Teff = 4635 $\pm$ 77~K, log g = 2.06 $\pm$ 0.20~dex, and [Fe/H] = -0.21$\pm$ 0.13~dex, HRV = --490.8 $\pm$ 0.8~km~s$^{-1}$) that are in very good agreement with those found in RAVE DR4. The confirmation of the impressively high metallicity for this HiVel star indicates that it is unlikely to have come from a dwarf galaxy that, on average, have significantly lower metallicities. The results of the abundance pattern in the $\alpha$-elements (Figure \ref{fig:alphachem}) and neutron-capture elements (Figure \ref{fig:schem}), of J2217 can be found in Figure \ref{fig:alpha+sprocess} and correspondingly in Table \ref{tab:highreschemJ22}. We also explored the Na-Ni abundances of J2217 in Figure \ref{fig:NaNi}. Studies as early as \cite{Nissen1997} indicated that [Na/Fe] and [Ni/Fe] may distinguish stars that were accreted from other field population stars. Exploring this relationship between Ni-Na will allow us to determine if J2217 is consistent with having being accreted. We found that the abundance pattern of J2217 namely the $\alpha$-elements, neutron-capture elements, and the abundance ratios of [O/Fe], [Ni/Fe], [Al/Fe], and [Na/Fe], most resembles the Galactic (thick) disk. It is worth mentioning that we found a relatively high enhancement in Barium, [Ba/Fe] = +0.35 dex, for J2217 however it is still within the range of the thick disk. Furthermore, the abundances in Table \ref{tab:highreschemJ22}, namely Fe, Na, and O, were compared with known globular clusters \citep[e.g.][]{Carretta2009} to determine if this star is consistent with a metal-rich globular cluster. This test indicated that J2217 is not consistent with any known globular clusters.

When we couple the abundance analysis with that of the orbital integration we favor a scenario in which this star was kicked out of the Galactic (thick) disk. The mechanism by which this star was ejected from the disk, namely dwarf galaxy heating \citep{Purcell2010}, binary supernova ejection \citep{Bromley2009} or other gravitational mechanisms, is unclear. Regardless, this provides observational support for the idea that the most metal-rich stars in the Galactic halo may have been assembled by kicking out Galactic disk stars \citep{Bromley2009}. 
\linebreak

\begin{table}
\caption{ARCES Elemental Abundances for J2217}
\label{tab:highreschemJ22}
\centering 
\begin{tabular}{c c c c c }
\hline\hline
Species & N & log $\epsilon$ (X) & $\sigma$ & {[}X/Fe{]} \\
\hline
Mg     & 8 & 7.56           & 0.14     & +0.25      \\
Ca   & 6 & 6.32           & 0.04     & +0.23      \\
Ti I   & 4 & 4.91           & 0.04     & +0.23      \\
Ti II     & 5 & 4.75           & 0.13     & +0.07      \\
Si    & 8 & 7.42              & 0.10        & +0.13          \\
C I    & 5 & 8.33              & 0.05        & +0.16          \\
O     & 1 & 8.69          & 0.10     &   +0.25 \\
Fe I   & 60 & 7.23           & 0.13     & ...      \\
Fe II   & 8 & 7.22           & 0.08     & ...      \\
Al     & 3 & 6.43           & 0.03     &   +0.28 \\
Na     & 5 & 617           & 0.09     &   +0.22 \\
Ni     & 5 & 6.15           & 0.09     &   +0.14 \\
Ba II     & 3 & 2.30           & 0.04    &   +0.35 \\
La II     & 6 & 0.93          & 0.06     &   +0.02 \\
Zr II     & 3 & 2.19           & 0.07     &   --0.18 \\
Sr I     & 1 & 2.73           & 0.10     &   +0.03 \\
Eu II    & 2 & 0.40           & 0.02     &   --0.19 \\
Cu I    & 3 & 3.78          & 0.10     &   --0.21 \\
\hline       
\end{tabular}
\\
\raggedright
NOTE: The chemical abundances of each species (column 1) are shown. Column 2 is the number of lines used to determine the abundance of each species. Column 3 is the log of the absolute abundance and the line-to-line dispersion is listed in column 4. Where there is only one line we quote a conservative error bar of $\pm$ 0.10 dex. Finally, column 5 is the solar relative abundances. 
\end{table}

\subsection{Hypervelocity Star Candidates}
HVSs are rare fast-moving stars which have velocities that exceed the Galactic escape speed \citep[$\sim$ 533 \kms at the solar circle,][]{Piffl2013}. Recent studies have used HVSs to better understand conditions and dynamics of the hidden Galactic Centre as well as the Galactic halo \citep{Kollmeier2010, Brown2009}.

We have identified one HVS candidate whose galactic rest-frame velocity is more than 1-$\sigma$ above the escape speed and three stars whose velocities, within the uncertainties, are above 500 \kms that are worth additional scrutiny. Their basic observational and kinematical properties can be found in Table \ref{tab:HVSobs} and Table \ref{tab:HVSkin}, respectively. J1610 has a metallicity of --0.86 dex and is at a Galactocentric radius of approximately 5 kpc and has a V$_{GRF}$ of 807 $\pm$ 154~km~s$^{-1}$. Assuming the escape speed of \cite{Piffl2013} at this radius ($\sim$ 600~km~s$^{-1}$), this candidate has a velocity at least 1-$\sigma$ above the escape speed and is among the first HVS candidates from RAVE. If the escape speed were significantly higher, above 650 \kms, this candidate would have a velocity less than 1-$\sigma$ escape speed but still impressively high. It is important to note there is relatively bright star close to this star and thus the uncertainties in the proper motion may be underestimated. More accurate distance and proper motion estimates will help decrease the error in order to confirm the total space velocity of this target.

To date, the tens of known HVSs are massive (3-4 M$_{\sun}$) early-type main sequence stars, including hot O, B, and A type stars \citep{Brown2009,Brown2012, Brown2014}. \cite{Li2012} and \cite{Palladino2014} have recently identified potential later-type HVS candidates which still need confirmation. The HVS candidate J1610 is among the few that are not early-type stars. Orbital integration of this HVS candidate indicates that it does not originate in the Galactic Centre. If confirmed, this would add to an increasing list of candidate HVS that do not seem to originate in the Galactic Centre \citep[e.g.][]{Palladino2014} and thus need an alternative production mechanism \citep[e.g.][]{Yu2003, Abadi2009, Herber2008, Przybilla2008}.

The second star, J1544, has a V$_{GRF}$ of 523 $\pm$ 40km~s$^{-1}$. This star has a velocity that is just below the escape speed at it position (which is expected to be $\sim$ 570 \kms). We have followed up J1544 with high-resolution echelle spectra from the ARCES instrument in order obtain detailed abundances in part because its high global metallicity ([M/H] = --0.54 dex). The spectrum of J1544 in the Mg I triplet region can be found in Figure \ref{fig:higresspec}. The stellar parameters of the star was determined using the high-resolution spectra and indicate that J1544 has a \Teff = 4458 $\pm$ 120 K; log g = 1.44 $\pm$ 0.2 dex; [Fe/H] = -1.24 $\pm$ 0.15 dex; $\xi = 1.77 \pm 0.06$ \kms which is consistent with the RAVE parameters. Furthermore, we have confirmed the very high heliocentric RV (--406.7 $\pm$ 0.80 km~s$^{-1}$) observed by RAVE for this star. Using the stellar parameters, we have determined the abundance of several $\alpha$ and neutron-capture elements. Table \ref{tab:highreschem} contains the high-resolution abundance analysis for J1544 including the species, number of lines used, log of the absolute abundance, the uncertainty (which was estimated as the standard deviation of the abundance determined from each individual line), and the solar-scaled abundance ratio. The [$\alpha$/Fe] for this star was determined by computing the mean abundance of the four $\alpha$-elements including Mg, Ca, Ti, and Si. 

\begin{table}
\caption{ARCES Elemental Abundances for J1544}
\centering 
\begin{tabular}{c c c c c }
\hline\hline
Species & N & log $\epsilon$ (X) & $\sigma$ & {[}X/Fe{]} \\
\hline
Mg    & 5 & 6.58           & 0.08     & +0.29      \\
Ca    & 8 & 5.30           & 0.13     & +0.23      \\
Ti I     & 5 & 3.76           & 0.09     & +0.10      \\
Ti II     & 1 & 3.74           & 0.10     & +0.08      \\
Si      & 7 & 6.50              & 0.06        & +0.23          \\
Al I     & 3 & 5.33           & 0.05     &   +0.20 \\
Na I     & 5 & 4.79           & 0.11     &   --0.14 \\
Ni I     & 5 & 4.90           & 0.10     &   --0.09 \\
Fe I     & 95 & 6.21           & 0.15     &   ... \\
Fe II     & 16 & 6.23           & 0.16     &   ... \\
Ba II     & 3 & 1.11           & 0.09     &   +0.16 \\
La II     & 5 & 0.13           & 0.04     &   +0.22 \\
Eu II    & 2 & -0.05           & 0.04     &   +0.65 \\
\hline       
\end{tabular}
\\
\label{tab:highreschem}
\raggedright
NOTE: The format is the same as in Table \ref{tab:highreschemJ22}.

\end{table}

The high-resolution abundances for the $\alpha$-elements (Figure \ref{fig:alphachem}) indicates that this star has a small enhancement with an [$\alpha$/Fe] = +0.21 $\pm$ 0.07 dex. At this metallicity, the star would be classified as one of the `low-$\alpha$' stars of \cite{Nissen2010} which they interpret as an `accreted' population. The neutron-capture elemental abundances of J1544 is consistent with both the field population and a dwarf galaxy like the Fornax (Figure \ref{fig:schem}). The Al and Mg ratios, [Al/Fe] = +0.20 $\pm$ 0.05 and [Mg/Fe] = +0.29 $\pm$ 0.08 dex respectively, are consistent with the field population \citep{Fulbright2002}. The abundance of [Na/Fe] and [Ni/Fe] are both depleted (Figure \ref{fig:NaNi}). Within the errors, the Na-Ni of this star is consistent with both the field star population of \cite{Venn2004} and \cite{Nissen2011} and massive dwarf galaxies. The chemical information from the high-resolution analysis indicates that this star is not consistent with the Galactic thin disk, because its metallicity is too low. The star also does not likely originate in the Galactic thick disk as one would expect a higher $\alpha$-abundance. We also compared the abundances of this star to known globular clusters with similar iron abundances \citep[e.g.][]{Carretta2009}. The depletion in Na and Ni combined with the abundances of Al and Mg do not seem consistent with globular clusters. Its chemistry suggests that this star may have a halo or dwarf galaxy formation site. 

It is also worth mentioning that in addition to the HVS candidate there are two other extremely high velocity bound stars whose total velocity vectors are at or near 500 km s$^{-1}$. Namely, J142103.5-374549 has a  v$_{GRF} =$ 460 $\pm$ 70 km s$^{-1}$ and J155304.7-060620 has a v$_{GRF} =$ 474 $\pm$ 43 km s$^{-1}$. Orbital integration shows J142103.5-374549 is on a highly elliptical orbit ($e$ = 0.91) getting as close as 3.72 kpc to the Galactic Centre. We note that J142103.5-374549 was excluded from \cite{Piffl2013} because of its unusual place on the Hertzsprung-Russell diagram. Further, both J142103.5-374549 and J155304.7-060620 have conflicting distance, and proper motions when comparing UCAC4 and distances from \cite{Binney2013} with those of \cite{Bilir2012} and \cite{Francis2013}.

\begin{table*}
\caption{Observational Properties of HVS and Metal-Rich HiVel Star Candidates} 
\centering 
\begin{tabular}{c c c c c c c c c c c} 
\hline\hline 
RAVEID&$\alpha$ & $\delta$&$\mu_{\alpha} cos(\delta)$&$\sigma\mu_{\alpha} cos(\delta)$&$\mu_{\delta}$&$\sigma\mu_{\delta}$&d$_{\sun}$&$\sigma$d$_{\sun}$ & [M/H] & [$\alpha$/Fe]\\

&($^{\circ}$) &($^{\circ}$)&(mas/yr)&(mas/yr)&(mas/yr)&(mas/yr)&(pc)&(pc)&(dex)&(dex) \\
\hline 
J155304.7-060620 &238.26967 &-6.10550 &-28.0 &1.3 &26.9 &1.3 &1480 &463 &-1.67 &0.33\\
 J142103.5-374549 &215.26442 &-37.76361 &8.7 &1.4 &3.1 &1.4 &3963 &2541 &-1.60 &0.50\\
 J154401.1-162451 &236.00438 &-16.41428 &2.6 &2.2 &3.8 &2.6 &3544 &1222 &-0.54 &0.80\\
 J161055.6-112009 &242.73183 &-11.33589 &-25.6 &2.2 &22.9 &2.4 &4369 &1451 &-0.82 &0.37\\
 J152905.9-365544 &232.27450 &-36.92883 &-4.8 &2.0 &3.4 &2.0 &2863 &738 &-0.21 &0.16\\
 J193647.0-590741 &294.19575 &-59.12814 &3.5 &1.1 &-8.4 &2.2 &618 &301 &-0.27 &...\\
 J221759.1-051149 &334.49604 &-5.19700 &-20.4 &1.1 &5.0 &1.1 &580 &147 &-0.18 &0.04\\
\hline 
\end{tabular} 
\label{tab:HVSobs} 
\end{table*} 

\begin{table*}
\caption{Kinematic Properties of HVS and Metal-Rich HiVel Star Candidates} 
\centering 
\begin{tabular}{c c c c c c c c c c c} 
\hline\hline 
RAVEID&HRV &$\sigma$HRV& U &$\sigma$U & V& $\sigma$V & W & $\sigma$W & V$_{GRF}$ & $\sigma$V$_{GRF}$ \\

&(\kms)&(\kms) &(\kms)&(\kms)&(\kms)&(\kms)&(\kms)&(\kms)&(\kms)&(\kms)\\
\hline 
J155304.7-060620 &-317.9 &0.9 &-408 &46 &235 &10 &47 &66 &473 &47 \\
 J142103.5-374549 &405.2 &1.1 &413 &76 &136 &96 &151 &29 &460 &73 \\
 J154401.1-162451 &-403.2 &1.7 &-335 &21 &358 &48 &-179 &37 &522 &29 \\
 J161055.6-112009 &-296.8 &1.0 &-589 &112 &249 &50 &492 &209 &807 &159 \\
 J152905.9-365544 &-189.8 &1.5 &-186 &15 &296 &24 &26 &33 &351 &17 \\
 J193647.0-590741 &-314.0 &1.8 &-261 &8 &317 &12 &146 &6 &436 &8 \\
 J221759.1-051149 &-491.0 &0.9 &-133 &10 &-23 &6 &404 &9 &426 &8 \\

\hline 
\end{tabular} 
\label{tab:HVSkin}
\end{table*} 

\section{Discussion and Conclusion}
\label{sec:conclusion}
We have aimed to characterize a set of HiVel stars in RAVE DR4. To do this, we applied a series of quality cuts on the initial RAVE DR4 ensuring the data have quality spectra and measurements needed to estimate the 6D position and velocity vector as well as estimates of metallicity. In order to maximize the HiVel stars in our sample, we selected stars which have absolute galactic-rest frame RVs (corrected for solar motion) $>$ 275 km s$^{-1}$. This led to some contamination from disk stars near $l$ = 90$^{\circ} $ and 270$^{\circ}$ where the primary component of the disk stars velocity is along the line-of-sight. To deselect these stars, we made a stricter cut in GRV within $ \pm$ 50$^{\circ}$ of the above Galactic latitudes. We sourced the distances from the estimated spectrophotometric parallaxes from \cite{Binney2013} and proper motions from the UCAC4 catalogue \citep{Zacharias2013} and combined it with the information from RAVE DR4 to obtain the full 6D position and velocity vectors. We also implemented an orbital integration code to determine orbital parameters, particularly Z$_{max}$ and eccentricity, to study the kinematics. We also studied the metallicity and [$\alpha$/Fe] abundances that compliment the kinematical study of our sample by attempting to measure the chemical distribution of our HiVel stars.

Our results can be summarized in the following way:
\begin{enumerate}
\item Kinematically, HiVel stars are mostly consistent with the Galactic halo (Figure \ref{fig:toomre}) and are characterized by eccentric orbits that can extend, on average, 14 kpc out of the Galactic plane. Chemically, HiVel stars in RAVE are metal-poor (peaking at [M/H] =  --1.2 dex) compared to the rest of RAVE (peaking [M/H] = --0.22 dex), which is consistent with the (inner) halo \citep{Kordopatis2013b, Carollo2007, Carollo2010, Piffl2013}. While the metal-weak thick disk overlaps the metallicity region of the HiVel stars, the rather hot kinematics that describe the HiVel stars (Figure \ref{fig:ezmax}) favors the Galactic halo as a current location. It is interesting to point out that the mean iron abundance of the inner Galactic halo from \cite{Carollo2007} is slightly more metal-poor compared to the mean \textit{global metallicity} of our HiVel stars.

\item The HiVel stars in our sample are, on average, $\alpha$-enhanced compared to the rest of the RAVE sample (Figure \ref{fig:alphadist}). The \afe distribution of the HiVel stars is consistent, within the 1-$\sigma$ error, to the $\alpha$-enhancement (\afe $\sim$ +0.4) expected of inner halo stars \citep{Haywood2013, Nissen2012, Nissen2010, Ruchti2010, Adibekyan2012, Sheffield2012}. The inner halo is thought to be formed of two components: an $\alpha$-rich component maybe formed in situ and an $\alpha$-poor component possibly accreted by dwarf galaxies \citep{Nissen2010, Schuster2012, Hawkins2014}. The large spread in $\alpha$-enrichment of the both RAVE ($\sim$ 0.18 dex) and HiVel stars ($\sim$ 0.25 dex) in our sample could be suggestive of an $\alpha$-rich and $\alpha$-poor population in both samples. However, it is more likely that this is blurred by the uncertainty in the [$\alpha$/Fe] estimates from the chemical pipeline of the RAVE DR4 \citep{Boeche2011, Kordopatis2013}. 

\item While most of the HiVel stars are metal-poor, there are several stars that have metallicity above --1.0 dex. These stars, while having kinematics that resemble halo stars, have disk like metallicity and thus do not conform to the rest of HiVel stars. One of these stars, J2217 has a particular high metallicity ([M/H] = --0.18). Depending on how one defines the metallicity distribution function in the halo this star is a $\sim$ 1.9-4$\sigma$ outlier \citep{An2013, Carollo2007, Carollo2010, Kordopatis2013b, Chiba2000}. Conversely, we can assess the probability this star being kinematically apart of the thin disk, thick disk, or halo. This star is described by an orbit with a Z$_{max}$ = 35 $\pm$ 10 kpc, overall eccentricity of $e$ = 0.72 $\pm$ 0.03, and a Galactic rest frame velocity of 426 $\pm$ 10 km~s$^{-1}$ (consistent with the Galactic halo).  Using the population classification from the full space motion described in \cite{Bensby2003}, we found this particular giant star is classified as a `high-probability' halo star. \cite{Ivezic2008} found that while the metallicity distribution of the inner halo peaks at -1.2 dex, the tail of the distribution extends up to approximately solar metallicities. Theoretically, \cite{Bromley2009} discusses the possibility of runaway disk stars being a contributor to the high-metallicity tail of the inner stellar halo. Could this star be observational evidence of this conjecture?

To help answer this, we obtained high-resolution spectrum J2217 (Figure \ref{fig:higresspec}) to do detailed chemical abundance analysis. We confirmed the stellar parameters of RAVE and measured the abundances of light elements, $\alpha$-elements and neutron-capture elements, which can be found in Table \ref{tab:highreschemJ22}. The abundances (see Figure \ref{fig:alpha+sprocess} and Figure \ref{fig:NaNi}) and the orbital integration (see Figure \ref{fig:orbitJ22175}) of J2217 confirm that this star was likely born in the Galactic thick disk. The abundances of Fe, Na, and O are not consistent with known globular clusters \citep[e.g.][]{Carretta2009}. Given the lack of s-process enhancement, or other chemical peculiarities, it is likely this star was kicked into the Galactic halo via gravitation mechanisms. Alternatively, given that nearly half of solar-type stars are in binary systems \citep{Duchene2013}, this star could also have been launched into the halo by a binary disruption event. The lack of neutron-capture and carbon enhancement however could point to the former scenario. The discovery of this runaway disk star indicates that while almost all HiVel stars currently reside in the Galactic halo, they were not necessarily \textit{born} in the halo.

\item We have found a HVS candidate using RAVE whose total Galactic rest frame velocity is larger the expected escape velocity (above the 1-$\sigma$ level). The HVS candidate is an evolved giant which is different than the known B-type HVS that are currently discussed in the literature \citep[e.g.][]{Brown2014,Brown2009}. 

\item We also followed up the star with the second highest velocity (with V$_{\mathrm{GRF}}$ larger than 500 \kms) in our sample (J1544). We confirmed the stellar parameters determined by RAVE of this star as well as measured abundance of several $\alpha$-elements, neutron-capture elements. The chemical abundances, found in Table \ref{tab:highreschem}, are consistent with either the halo field star origin or (massive) dwarf galaxy origin (see Figure \ref{fig:alpha+sprocess} and Figure \ref{fig:NaNi}). The high speed, total Galactic rest frame velocity of 526 $\pm$ 40 \kms, is near but not above the Milky Way escape speed at its position and is unusually fast for a halo field star. Furthermore, the orbital integration indicates this object is currently passing through the disk (Figure \ref{fig:orbitJ22175}). While it may be possible for this star to be born in the halo and achieve such a high velocity it is also possible that the star has been accreted from a dwarf galaxy \citep[e.g.][]{Abadi2009, Piffl2011} and may explain its velocity near the escape speed. The chemodynamics of this particular star is consistent with either scenario leaving its origins unknown. 

\end{enumerate}

It is interesting to study the chemistry and kinematics together of the HVS and HiVel star candidates as they give us invaluable information about the formation environment of these populations. Our analysis has shown that HVS that are near the Galactic escape speed should have complimenting chemical information to better constrain its formation environment. \cite{Palladino2014} has found a set of 20 metal-poor HVS candidates with SDSS of those 6 have velocities within 100 km s$^{-1}$ of the escape speed assuming spherical potential. Could these be captured stars as well? Complementary chemical abundance analysis may help decipher where these stars originate from and thus shed light on potential formation mechanisms for HVS. With the upcoming Gaia mission, we will better be able to constrain the distances, proper motions and thus the total velocity vector for all of our stars but more specifically the HVS candidates.

\section*{Acknowledgements}
We would like to thank the anonymous referee whose comments improved this manuscript. We also would like to thank P. Jofr\'e, A. Casey, and V. Belokurov for discussions that greatly improved this work. K.H. is funded by the British Marshall Scholarship program and the King's College, Cambridge Studentship. R. F. G. W. acknowledges funding from the NSF grant OIA-1124403. 
This work is based on observations obtained with the Apache Point Observatory 3.5-meter telescope, which is owned and operated by the Astrophysical Research Consortium.
Funding for RAVE has been provided by: the Australian Astronomical Observatory; the Leibniz-Institut fuer Astrophysik Potsdam (AIP); the Australian National University; the Australian Research Council; the French National Research Agency; the German Research Foundation (SPP 1177 and SFB 881); the European Research Council (ERC-StG 240271 Galactica); the Istituto Nazionale di Astrofisica at Padova; The Johns Hopkins University; the National Science Foundation of the USA (AST-0908326); the W. M. Keck foundation; the Macquarie University; the Netherlands Research School for Astronomy; the Natural Sciences and Engineering Research Council of Canada; the Slovenian Research Agency; the Swiss National Science Foundation; the Science \& Technology Facilities Council of the UK; Opticon; Strasbourg Observatory; and the Universities of Groningen, Heidelberg and Sydney. The RAVE web site is at http://www.rave-survey.org 

\bibliography{Hawkins_Hivel.bib}
\label{lastpage}

\end{document}